**Axon Membrane Skeleton Structure is Optimized for Coordinated Sodium Propagation**

**Authors**

Yihao Zhang[#,1], Vi Ha[#,1], He Li[#,2], Anastasios V. Tzingounis[3], George Lykotrafitis[1,4,*]

[1] Department of Mechanical Engineering, University of Connecticut, Storrs, CT 06269, USA

[2] Division of Applied Mathematics, Brown University, RI 02912, USA

[3] Department of Physiology and Neurobiology, University of Connecticut, Storrs, CT 06269, USA

[4] Department of Biomedical Engineering, University of Connecticut, Storrs, CT 06269, USA

[#] These authors contributed equally to this work

**\*Correspondence**

george.lykotrafitis@uconn.edu

**SUMMARY**

Axons transmit action potentials with high fidelity and minimal jitter. This unique capability is likely the result of the spatiotemporal arrangement of sodium channels along the axon. Super-resolution microscopy recently revealed that the axon membrane skeleton is structured as a series of actin rings connected by spectrin filaments that are held under entropic tension. Sodium channels also exhibit a periodic distribution pattern, as they bind to ankyrin G, which associates with spectrin. Here, we elucidate the relationship between the axon membrane skeleton structure and the function of the axon. By combining cytoskeletal dynamics and continuum diffusion modeling, we show that spectrin filaments under tension minimize the thermal fluctuations of sodium channels and prevent overlap of neighboring channel trajectories. Importantly, this axon skeletal arrangement allows for a highly reproducible band-like activation of sodium channels leading to coordinated sodium propagation along the axon.

**INTRODUCTION**

It is known for some time that microtubules and neurofilaments are the predominant structural filamentous proteins in the axon (Hirokawa, 1982; Yu and Baas, 1994). However, the structure of the axon plasma membrane skeleton was only discovered very recently (Xu et al., 2013). Super-resolution fluorescence microscopy (Rust et al., 2006) revealed that the membrane skeleton of an unmyelinated axon consists of actin filaments, capped with adducin at one end, that form ring-like structures along the circumference of the axon. The actin rings are connected via spectrin tetramers oriented along the longitudinal direction of the axon (Figure 1A). This cytoskeletal structure is extended across the entire axon. The distance between the periodic actin rings is approximately 180 to 190 nm (Lukinavicius et al., 2014; Xu et al., 2013; Zhong et al., 2014). Each spectrin tetramer is formed by the association of two identical heterodimers comprising an α-chain and a β-chain with 22 and 17-triple-helical segments, respectively (Bennett and Baines, 2001). In the distal axon, each heterodimer consists of two intertwined αII- and βII-spectrin chains running antiparallel to one another (Figure S1A). The axon initial segment (AIS) of mature neurons, however, appears to contain the subtype β-IV spectrin instead of β-II spectrin (Zhong et al., 2014).

Spectrin tetramers are associated with the lipid bilayer (Xu et al., 2013) in a manner similar to that occurring at the red blood cell (RBC) membrane (Bennett and Baines, 2001; Mohandas and Evans, 1994) (Figure S1B). In RBCs, ankyrin plays a major role in anchoring the lipid bilayer to the spectrin network by associating a spectrin filament with the anion exchanger integral membrane protein band-3 (Lux, 1979; Tse and Lux, 1999) (Figure S1A). Ankyrin binds to the middle of the spectrin tetramer, at the $15^{th}$ repeat of β-spectrin near its carboxyl terminus (Bennett and Baines, 2001; Bennett and Healy, 2009), and at the same time it binds to the cytoplasmic domain of band-3 (Lux, 1979). In axons, the spatial distribution of ankyrin-G and ankyrin-B is highly periodic in the proximal and distal area of the axon, respectively (Xu et al., 2013; Zhong et al., 2014). Because ankyrin binds to several molecules it plays a very significant role in the organization of the axonal plasma membrane, in addition to anchoring the spectrin network to the lipid bilayer. Voltage-gated Na ($Na_v$) channels can bind to subdomains 3 and 4 of ankyrin (Srinivasan et al., 1992). Since ankyrin appears in a periodic pattern in the axon, the $Na_v$ channels exhibit a periodic distribution pattern in the AIS alternating with the N terminus of βIV-spectrin (Xu et al., 2013).

The periodic structure of the membrane skeleton is thought to convey structural durability to the axon. The principal argument to support this conjecture comes from the similarity between the structural elements of the plasma membrane skeletons of the RBC and of the neuronal axon. However, a simple examination of the two structures shows that their geometrical arrangements are radically different resulting in hyperelastic flexibility for the RBC cytoskeleton and in reduced radial deformability and longitudinal extensibility for the axon. The two most important differences between the geometric configurations of the RBC and of the axon membrane skeletons are the following. First, the RBC cytoskeleton forms an approximately six-fold symmetric two-dimensional network that behaves as an incompressible hyperelastic material (Dao et al., 2003). In the axon, actin filaments form rings along the circumference, connected by spectrin filaments oriented along the axon. In this case, the membrane skeleton assumes the form of a two-dimensional cylindrically symmetric orthotropic network (Figure 1). The resistance from the membrane skeleton to compression along the axon is orders of magnitude smaller than the resistance to radial compression. The second important difference is that in the case of the RBC membrane, the end-to-end distance of the spectrin tetramers is $\sim 75 nm$, which is close to the end-to-end distance of a free spectrin filament (Stein and Bronner, 1989). This suggests that the spectrin network in the RBC is near equilibrium. In the axon, however, the distance between the actin rings was reported to be approximately 180 to $190 nm$ (Lukinavicius et al., 2014; Xu et al., 2013). Thus, the end-to-end distance of the spectrin tetramers is 145 to 155 nm, which is close to their contour length of approximately $200 nm$ (Glenney Jr et al., 1982; Shotton et al., 1979). This means that the spectrin filaments in the axon membrane skeleton are held under entropic tension and the flexibility of the network along the axon might be limited.

Most of the discussion concerning the structure of the axon skeleton has been focused on questions related to the mechanical stability of the axon. However it is also necessary to examine the role of the axon skeleton with respect to the main function of the axon, the propagation of the action potential. Consequently, in this work, we explore the role of the membrane axon skeleton in relation to sodium dynamics. By combining coarse-grain molecular dynamics and diffusion modeling we show that the axon membrane skeleton structure is to facilitate coordinated and highly reproducible propagation of sodium signals down the axon.

**RESULTS AND DISCUSSION**

We developed a basic axon membrane skeleton dynamics model reflecting the recent super-resolution studies (Figure 1B; for details see Experimental Procedures). The model represents actin rings encircling the axon connected by spectrin tetramers that are oriented along the longitudinal direction of the axon. We employed two types of finitely extendable nonlinear elastic potentials to stabilize the circular shape of the actin rings and the distance between each consecutive actin ring. We represented each spectrin tetramer as a single chain of spherical particles connected via proper harmonic and repulsive Lennard-Jones potentials (see Experimental Procedures and Figure S2). Ankyrin proteins, depicted as green particles in Figure 1B, are connected to spectrin by a harmonic potential. Finally, we used a harmonic potential to confine the radial motion of spectrin and ankyrin particles to the lipid bilayer rather than explicitly representing the lipid bilayer. The persistence length ($l_p$) of a spectrin filament is typically between $10nm$ and $20nm$ (Stokke et al., 1985; Svoboda et al., 1992) and its contour length ($L_c$) is approximately $200nm$. Thus, the end-to-end distance of a spectrin filament at equilibrium ranges from $63nm$ to $89.5nm$, based on the equation $\langle \mathbf{r}_{ee}^2 \rangle^{1/2} \cong \sqrt{2 l_p L_c}$. Spectrin tetramers of the axon membrane skeleton have an end-to-end distance of approximately $150 nm$; therefore, they are under tension with a reduced range of thermal motion. To demonstrate this, we equilibrated the model for $15 \times 10^4$ time steps at constant volume and temperature (Figure 2A) and recorded the thermal motion of ankyrin particles for $5 \times 10^6$ time steps. The trajectory of ankyrin particles outlined an area with an average radius of $\sim 5.37 nm$. As Figure 2A shows, the ankyrin particles and consequently the connected $Na_v$ channels maintained an ordered configuration, in contrast to simulations with spectrin not under tension (Figure 2B; equilibrium end-to-end distance of $75 nm$).

To examine the functional significance of the axon structure, we established a 3-D continuum diffusion space model comprised of isotropically and homogeneously arranged nodes that interface the Cartesian space of the axon membrane skeleton (see Experimental Procedures). We represented the axon as a cylindrical tube and positioned the $Na_v$ channels at the nodes of the continuum space. For each $Na_v$ we chose the node which was closest to a random location in the area delineated by the corresponding ankyrin particle during its thermal motion. The random location was generated using a Gaussian random number generator with standard deviation equal to the radius of the area outlined by the corresponding ankyrin particle. Nodes with a distance from the longitudinal axis greater than the radius of the actin rings were treated as "insulators" (there was no ion flux). Nodes with a distance from the longitudinal axis less than $1/4$ of the axon radius were treated as 'sinks' (the ion concentration remained zero). The proposed configuration gives a $Na_v$ channel density of ~150 channels per $\mu m^2$, which is within the reported range of 110 to 300 channels per $\mu m^2$ in the axon initial segment (Kole and Stuart, 2012). We considered the node corresponding to the $Na_v$ channel to be an ion source activated when the local sodium concentration reached a pre-determined value, generating an inward flux of ions perpendicular to the axon surface.

To analyze the effects of the axon membrane skeleton structure in respect to sodium propagation, we first set the distance between the actin rings to $185 nm$ in accordance with experimental data (Lukinavicius et al., 2014; Xu et al., 2013). The initial synchronous activation of the $Na_v$ channels led to a band of sodium influx in the axon consistent with the spatial ring-like arrangement of $Na_v$ channels (Figure 3A). As sodium diffused away and laterally to the membrane, the increasing sodium concentration brought the neighboring actin rings to threshold, causing another sodium band-like activation. This activity proceeded continuously down the axon leading to a coordinated propagation of sodium ion flux with high fidelity and reproducibility (Figure 3B).

We speculate that this unremitting sodium propagation is a result of the tension of the spectrin tetramers, which constrains their thermal fluctuation and consequently the thermal motion of ankyrin proteins and $Na_v$ channels. To test this hypothesis, we recorded the motion of ankyrin particles when the distance between consecutive actin rings was $110 nm$, corresponding to an end-to-end distance of $\sim 75 nm$ for the spectrin tetramers (Figure 2B). Importantly, this distance is close to their equilibrium distance meaning that the spectrin tetramers are not under tension but rather they have a slack. Using this model, we found that the thermal fluctuation of ankyrin particles was larger, covering circular areas with an average radius of $\sim 10.6 nm$. The ratio between this average radius and the end-to-end average distance of spectrin filaments was $\sim 0.141$ in contrast to simulations with spectrin under tension where the ratio was $\sim 0.036$. In addition, the areas between neighboring ankyrin particles overlapped (Figure 2B). Both changes led to a propagating sodium whirlpool rather than a coordinated sodium ring down the axon. This propagation caused asynchronous sodium channel activation with increased jitter and no reproducibility along the axon (Figures 3C and 3D).

Next, we investigated whether the circumferential distance between $Na_v$ channels is also critical for the propagation of sodium along the axon when spectrin filaments are under tension or with a slack. To test this, we decreased the number of channels per ring from 39 to 13 by alternating $Na_v$ channels in the continuum diffusion space with ankyrin particles. With this arrangement, the propagation of the sodium concentration spike was still unstable at a 110$nm$ ring-to-ring distance when spectrin filaments were not under tension but to a lower degree than the higher density of $Na_v$ channels (Figure 4). This result suggests that the circumferential distance between $Na_v$ channels interferes with sodium concentration peak formation and propagation, and that the axon membrane skeleton is important to maintain a high density of sodium channels along the circumference of the axon.

In sum, our membrane skeleton dynamics and continuum diffusion models revealed that the spatial distribution of $Na_v$ channels imposed by the subcellular structure of the axon is optimized for proper coordinated propagation of the sodium signal down the axon.

**MODEL DESCRIPTION**

**Axon membrane skeleton model**

The proposed model reflects the structure of the membrane skeleton of the proximal and distal unmyelinated axon as previously described (Xu et al., 2013; Zhong et al., 2014). The model is a representation of actin rings, oriented along the circumference of the axon, that are connected by spectrin tetramers tethered to the lipid bilayer at their middle section (Figure 1).

*Modeling spectrin tetramers*

A spectrin tetramer consists of two identical, intertwined, head-to-head associated heterodimers (Shotton et al., 1979). Each heterodimer is comprised of an α-spectrin and a β-spectrin chain consisting of 22 and 19 homologous triple helical repeats, respectively (Bennett and Baines, 2001). In our model, a spectrin tetramer is represented as a single chain of 41 spherical beads (gray particles in Figure S2A) connected by 40 harmonic springs. The solid red line in Figure S2B reflects the spring potential, $U^{SS}(r) = 1/2\, k_0 (r - r_{eq}^{SS})^2$, where $r$ is the distance between two consecutive spectrin particles, $r_{eq}^{SS} = L_c/40 = 5\,nm$ is the equilibrium distance between the spectrin particles (close to the size of the spectrin repeats), $L_c \simeq 200\,nm$ is the contour length of the spectrin tetramers, and $k_0$ is the spring constant (defined below). All spectrin particles interact via the repulsive Lennard-Jones (L-J) potential:

$$U_{rep}(r_{ij}) = \begin{cases} 4\varepsilon \left[ \left(\dfrac{\sigma}{r_{ij}}\right)^{12} - \left(\dfrac{\sigma}{r_{ij}}\right)^{6} \right] + \varepsilon & r_{ij} < R_{cut,LJ} = r_{eq}^{SS} \\ 0 & r_{ij} > R_{cut,LJ} = r_{eq}^{SS} \end{cases}, \qquad (1)$$

where $\varepsilon$ is the energy unit, $\sigma$ is the length unit, and $r_{ij}$ is the distance between spectrin particles. Setting the diameter of the spectrin particles equal to the equilibrium distance of the L-J potential $r_{eq}^{SS} = 2^{1/6}\sigma = 5nm$, the length scale is $\sigma = 4.45nm$. We chose the cutoff distance of the potential $R_{cut,LJ}$ to be the equilibrium distance $r_{eq}^{SS}$ between two spectrin particles. The potential is plotted as a dashed red line in Figure S2B. We made the spring constant $k_0 = 57\ \varepsilon/\sigma^2$ identical to the curvature of $U_{LJ}(r_{ij}) = 4\varepsilon\left[\left(\sigma/r_{ij}\right)^{12} - \left(\sigma/r_{ij}\right)^{6}\right]$ at equilibrium to reduce the number of free parameters (Li et al., 2007).

We performed a molecular dynamics simulation to compute the end-to-end distance $\langle r_{ee}^2 \rangle^{1/2}$ of the spectrin chain model at the temperature $T = 0.22\varepsilon/K_B$ for which the membrane skeleton is equilibrated, where $K_B$ is Boltzmann's constant. We first equilibrated the filament for $10^5$ time steps, and then measured the end-to-end distance for $3\times 10^6$ time steps during its thermal fluctuations. The recorded distances follow a Gaussian distribution $P(r_{ee}) = 1/\left(\lambda\sqrt{2\pi}\right)\exp\left[-\left(r_{ee} - \langle r_{ee}\rangle\right)^2/2\lambda^2\right]$, where $\lambda = \sqrt{\langle\left(r_{ee} - \langle r_{ee}\rangle\right)^2\rangle}$, and with a mean value of $\langle r_{ee}^2 \rangle^{1/2} = 79nm$ (Figure S3). For flexible filaments with $l_p << L_c$, the end-to-end distance is correlated with persistence length and contour length via the expression $\langle r_{ee}^2 \rangle^{1/2} \cong \sqrt{2l_p L_c}$. Taking into consideration that the spectrin contour length is approximately 200 nm (Glenney Jr et al., 1982; Shotton et al., 1979), we calculated the persistence length to be $15.6nm$. This result is close to experimentally reported values of approximately $20nm$ (Stokke et al., 1985) and $10nm$ (Svoboda et al., 1992).

*Modeling the actin rings*

The actin rings consist of short actin filaments capped by adducin along the circumference of the axon (Xu et al., 2013). In the particle model, an actin ring is represented as a collection of 39 beads (red particles in Figures 1B and insert and in Figure S2A) with a diameter of approximately $35nm$. These beads form a circle with a diameter of approximately $434nm$, which lies within the range of experimental results (Xu et al., 2013; Zhong et al., 2014). We chose the diameter of the actin particles to be $35nm$ based on values for the RBC membrane skeleton, which comprises short actin protofilaments (consisting of approximately 13 to 15 subunits) with a length of $33\pm 5nm$ (Fowler, 1996; Liu et al., 1987; Shen et al., 1986).

Two adjacent actin particles in the same ring connect via a spring potential $U^{AA} = 1/2 k_0 \left( r - r_{eq}^{AA} \right)^2$, with equilibrium distance $r_{eq}^{AA} = 35nm$, and a repulsive L-J potential, with $R_{cut,LJ} = r_{eq}^{AA}$ (shown as purple lines in Figure S2B). We employed a finitely deformable nonlinear bending potential that behaves as a finitely extendable nonlinear elastic (FENE) potential to maintain the circular shape of the actin rings. The potential has the form $U_b = -\frac{1}{2} k_b \Delta\theta_{max} \ln\left[ 1 - \left( \frac{\theta - \theta_0}{\Delta\theta_{max}} \right)^2 \right]$, where $k_b = 8\times 10^3 K_B T$ is the parameter that directly regulates the bending stiffness of the actin filament, and $\theta$ is the angle formed by three consecutive particles of the same ring. $\theta_0 = \frac{180°(39-2)}{39} = 170.77°$ is the equilibrium angle and $\Delta\theta_{max} = 0.3\theta_0$ is the maximum allowed bending angle. The employed value of $k_b$ produces a bending rigidity $\kappa = 5.3\times10^{-25} Nm^2$ for a straight stiff filament based on numerical calculations shown in (Li et al., 2012), which is approximately one order of magnitude larger than the bending rigidity of actin filaments (Gittes et al., 1993; Isambert et al., 1995). We opted to use this value because there is no experimental value for the actin rings of the axon and there is a possibility that the bending resistance of the rings is enhanced by associated microtubules. Regardless, the chosen value for $k_b$ guaranties that the shape of the rings is stable, and it does not affect the overall conclusions of the paper.

*Modeling the axon membrane network*

To build a mechanically stable network, we first connected each spectrin filament at its two ends to actin particles belonging to consecutive actin rings via an unbreakable linear spring potential $U^{AS}(r_{ij}) = 1/2 k_0 (r_{ij} - r_{eq}^{AS})^2$, where $r_{ij}$ is the distance between actin and spectrin particles and $r_{eq}^{AS} = 20nm$ is the equilibrium distance, and a repulsive L-J potential, with $R_{cut,LJ} = r_{eq}^{AS}$ (blue lines in Figure S2B). Second, microtubules are thought to play an important role in maintaining the polarity and structure of the AIS through interactions with the axonal cytoskeleton. In this model, the effect of microtubules on the structural integrity of the axon was implemented implicitly. We considered that microtubules interact with actin to maintain the equilibrium distance of actin rings at 185nm. To achieve this, we applied the FENE potential

$$U_{mt} = -\frac{1}{2} k_{mt} \Delta d_{max} \ln\left[1 - \left(\frac{d - d_{eq}^{RR}}{\Delta d_{max}}\right)^2\right]$$

on all actin particles belonging to consecutive rings. $k_{mt}$ is the parameter that determines the stiffness of the nonlinear spring between two actin rings, $d$ and $d_{eq}^{RR} = 185 nm$ are the distance and the equilibrium distance between the centers of the two actin rings, respectively (Xu et al., 2013; Zhong et al., 2014), and $\Delta d_{max} = 0.3 d_{eq}^{RR}$ is the maximum allowed deformation. The position of the center of each ring is calculated by utilizing the mean value of the $z$ coordinate of the actin particles. The choice of $k_{mt}$ is justified based on the following rationale: For small deformations, the FENE potential is approximated by $U_{mt} = \frac{1}{2}(k_{mt}/\Delta d_{max})(d - d_{eq}^{RR})^2$, which corresponds to a harmonic potential with a spring constant $K_{sp} = k_{mt}/\Delta d_{max}$. In this case, we can assume that $\tau = Eh$, where $\tau$ is the stress, $E$ is the Young's modulus of the axon, and $h$ is the strain. The final equation is $F/A = E(\Delta L/L)$, where $F = k_{sp}^t \Delta L$ is the force applied on the cross-section of the axon $A = \pi R^2$, where $k_{sp}^t = k_{sp}/(N-1)$ is the spring constant for the total axon, $R = 217 nm$ is the radius of the axon, $\Delta L$ is the elongation of the axon, $L = (N-1)d_{eq}^{RR}$ is the length of the axis, and $N$ is the number of springs. Combining the equations above, we determined that $k_{sp} = E\pi R^2/d_{eq}^{RR}$ and consequently $k_{mt} = k_{sp} \Delta d_{max} = 0.3 E\pi R^2$. A reasonable value for the axon's Young's modulus is $E \simeq 10 KPa$ (Javid et al., 2014), resulting in $k_{mt} \simeq 477 K_B T/\sigma \simeq 19,822 K_B T/d_{eq}^{RR}$, at $T = 300\,^oK$.

The final aspect of the model is the association between the axon membrane skeleton and the lipid bilayer. In RBCs, the membrane skeleton is anchored to the lipid bilayer via glycophorin at the actin junction complexes and via the integral membrane protein band-3 and ankyrin at the middle section of spectrin tetramers (Lux, 1979; Tse and Lux, 1999) (Figure S1). Regarding the association between a spectrin filament and the lipid bilayer in RBCs, ankyrin binds at the 15[th] repeat of β-spectrin near its carboxyl terminus, at the middle section of the spectrin tetramer (Bennett and Baines, 2001). At the same time, it binds to the cytoplasmic domain of band-3 (Lux, 1979), mediating the anchoring of spectrin filaments to the lipid bilayer. In the case of the neuronal axon, we considered the following experimental findings: (i) The spatial distribution of ankyrin-G is highly periodic in the proximal area of the axon, while ankyrin-B also exhibits a periodic pattern in distal axons (Xu et al., 2013; Zhong et al., 2014), (ii) $Na_v$ channels exhibit a periodic distribution pattern in the AIS alternating with actin rings (Xu et al., 2013), (iii) $Na_v$ can bind to subdomains 3 and 4 of ankyrin (Srinivasan et al., 1992), and (iv) Ankyrin-G and sodium channels are in 1:1 molar ratio in the brain. Based on these findings and on the fact that ankyrin binds near the carboxyl terminus of β-spectrin it is reasonable to assume that $Na_v$ channels are arranged in a periodic pattern along the axon via their association with ankyrin in a manner similar to band-3 association with spectrin in the RBC membrane. We also note that by assigning one $Na_v$ channel per ankyrin molecule, and consequently per spectrin tetramer, the $Na_v$ channel density is approximately 150 channels per $\mu m^2$, which lies within the range of 110 to 300 channels per $\mu m^2$ in AIS (Kole and Stuart, 2012).

To represent the anchoring of spectrin tetramers to the lipid bilayer, we used the following approach: We connected an ankyrin particle (depicted as a green particle in Figures 1B and Figure S2A) to the 20$^{th}$ particle of the spectrin filament by the spring potential $U^{SK}(r_{ij}) = 1/2 k_0 (r_{ij} - r_{eq}^{SK})^2$, where the radial equilibrium distance is $r_{eq}^{SK} = 15 nm$, (black solid line in Figure S2B). This distance corresponds to the radius of a spectrin particle ($2.5 nm$) and the effective radius of the cytoplasmic domain of the ankyrin complex connected to an $Na_v$ channel ($\sim 12.5 nm$) (Bamberg and Passow, 1992). We also implemented a repulsive L-J potential, with $R_{cut,LJ} = r_{eq}^{SK}$ (dashed black line in Figure S2B). For simplicity, we did not use a representation of the lipid bilayer in this model. Instead, we used a spring potential to represent the confinement applied on the motion of ankyrin particles and spectrin filaments by the lipid bilayer. The harmonic potential is given by $U^{LB}(r) = 1/2 k_c (r - r_0)^2$, where $r$ is the radial distance of spectrin and ankyrin particles from the central axis of the axon, and $r_0$ is the equilibrium distance from the central axis. We considered $r_0$ to be $217 nm$ and $232 nm$ for the spectrin and actin particles, respectively. Because the ankyrin particles are attached to the bilayer, the confinement potential acts on both radial directions (inwards and outwards). However, only the outward motion of the spectrin particles is confined, since the spectrin filament cannot cross the lipid bilayer. In contrast, the inward motion will not face additional constrain. The confinement stiffness in this model is $k_c = 0.1 k_0$.

*Membrane skeleton dynamics simulation details*

The configuration used in this paper consists of N = 16,029 particles, corresponding to an axon length of approximately $1.85 \mu m$. The numerical integrations of the equations of motion are performed using the Beeman algorithm. The temperature of the system is maintained at $K_B T / \varepsilon = 0.22$ by employing the Berendsen's thermostat (Berendsen et al., 1984), where $K_B$ is Boltzmann's constant and $T$ is the temperature. The model is implemented in the *NVT* ensemble (constant number of particles *N*, constant volume *V*, and constant temperature *T*). The time scale is $t_s = \sqrt{m\sigma^2/\varepsilon}$, the time step is $dt = 0.01 t_s$, and $m$ is the unit mass of the spectrin particles. We selected the temperature to render the conformation time of the spectrin filaments close to expected theoretical values (Li and Lykotrafitis, 2014). We gradually brought the model to the equilibrium length and temperature, and then equilibrated it for $15 \times 10^4$ time steps. We performed the measurements during a period of $5 \times 10^6$ time steps after equilibration.

**Diffusion model**

Our diffusion model describes the propagation of a sodium ion concentration spike along the neuronal axon. Diffusion occurs in a 3D Cartesian space for which the $z-$axis is identical to the $z-$axis (axon longitudinal axis) of the overlapping space in which the membrane skeleton particles move. The diffusion space comprises $41 \times 41 \times 600$ nodes. The axon is represented as a cylindrical tube with the *z*-axis corresponding to the longitudinal axis of the cylinder. All nodes with distance from the longitudinal axis greater than the radius of the actin rings were considered 'outside' of the axon. These were treated as insulators, meaning that there was no ion flux across the boundary, employing $(\vec{\nabla} C) \cdot \hat{\mathbf{n}} = 0$, where $C$ is the concentration and $\hat{\mathbf{n}}$ is the unit vector normal to the cylindrical surface of the axon. Nodes with a distance from the longitudinal axis of less than $1/4$ of the axon radius were treated as 'sinks', meaning that the ion concentration at these nodes remained zero at all times, using $(C(r \leq R/4, t) = 0)$, where $R$ is the radius of the axon.

To position the Na$_v$ channels in the diffusion space, we chose the nodes which were closest to randomly generated locations within the areas outlined by the corresponding ankyrin particles during their thermal motion. The random locations ware generated using a Gaussian random number generator with standard deviation equal to the average radius of the areas described by the ankyrin particles.

To determine the propagation of ion concentration along the axon, we solved the parabolic equation

$$\frac{\partial C}{\partial t} = D\left(\frac{\partial^2 C}{\partial x^2} + \frac{\partial^2 C}{\partial y^2} + \frac{\partial^2 C}{\partial z^2}\right), \tag{2}$$

where $D$ is the diffusion coefficient and $t$ is time.

The boundary conditions were defined as

$$\left(\vec{\nabla} C(r \geq R, t)\right) \cdot \hat{\mathbf{n}} = 0, \tag{3a}$$

and

$$C(r \leq R/4, t) = 0, \tag{3b}$$

The initial conditions were defined at $t = 0$ as $C_0 = 1$ for all channels of the first ring and $C = 0$ for all other nodes.

During the simulation, we set $C = C_0$ for five time steps at a node representing a Na$_v$ channel when the concentration at the node was higher than the threshold $(C \geq 10^{-4} C_0)$.

To solve the equation, we employed a forward-time central-spaced scheme

$$C_{i,j,k}^{l+1} = C_{i,j,k}^{l} + \Delta t \cdot D \left( \frac{C_{i+1,j,k}^{l} - 2C_{i,j,k}^{l} + C_{i-1,j,k}^{l}}{\Delta x^2} + \frac{C_{i,j+1,k}^{l} - 2C_{i,j,k}^{l} + C_{i,j-1,k}^{l}}{\Delta y^2} + \frac{C_{i,j,k+1}^{l} - 2C_{i,j,k}^{l} + C_{i,j,k-1}^{l}}{\Delta z^2} \right)$$

where $C_{i,j,k}^{l}$ is the concentration at the node with indices $i, j,$ and $k$, the superscript index $l$ determines the time step, and $\Delta x = \Delta y = \Delta z \simeq 2.7\sigma$ are the distances between the nodes in the $x$, $y$, and $z$ directions, respectively. The time step $\Delta t$ was determined by the equation $\Delta t = \Delta x^2 / (6D)$ to ensure the stability of the algorithm. The diffusion coefficient was chosen to be $D = 0.5(2.7\sigma)^2 / t_s$. Because the diffusion coefficient for sodium ions is $D \simeq 1.3 \times 10^{-9} \, m^2/s$ (Fell and Hutchison, 1971; Samson et al., 2003) and $\sigma = 4.45 nm$, the time step of our simulations is $\Delta t = 1.8 \times 10^{-8} s$. This time step is close to the time step obtained in RBC membrane simulations (Li and Lykotrafitis, 2014). Thus, the combination of skeleton dynamics simulations and propagation of ion concentration spike is compatible in terms of time scale.

**SUPPLEMENTAL INFORMATION**

Supplemental Information includes 3 figures and can be found attached to this article.

**AUTHOR CONTRIBUTIONS**

Y.Z, V.H., H.L., and G.L. designed simulations, carried out simulations. Y.Z., A.V.T., and G.L. analyzed the data and wrote the manuscript.

**ACKNOWLEDGMENTS**

This work was supported by a National Science Foundation (CMMI-1354363, NSF CARRER) grant awarded to G.L. and a National Institutes of Health (NIH/NINDS (NS073981) R01 grant awarded to A.V.T.

**FIGURES**

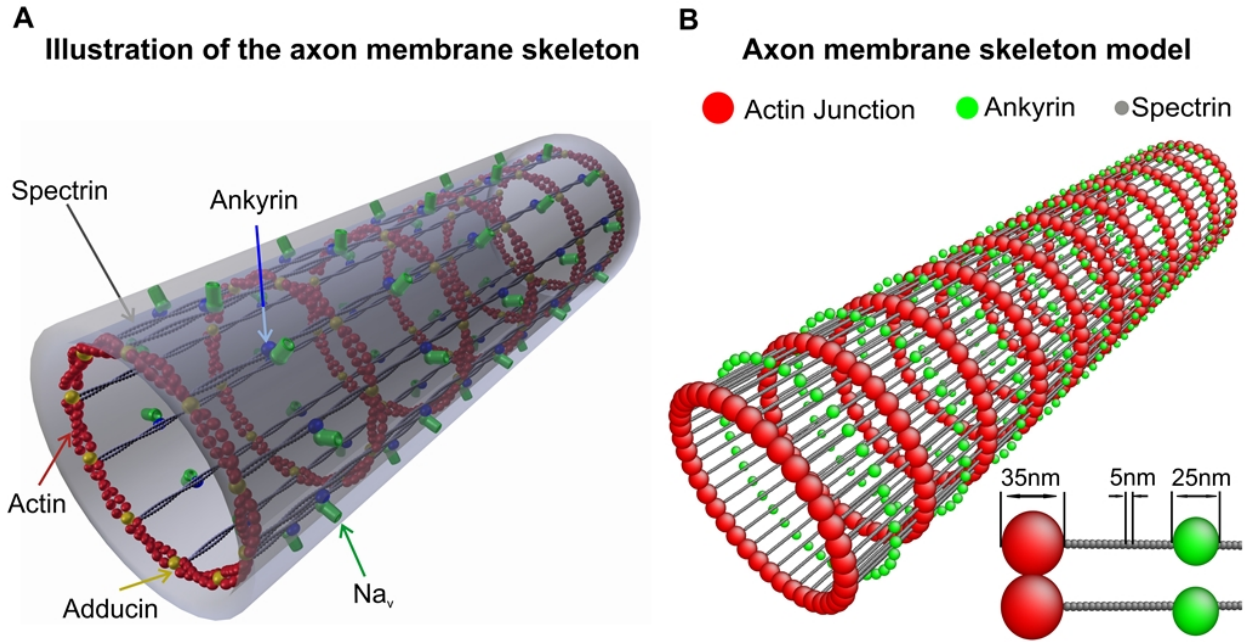

**Figure 1.** (A) Illustration of the axon membrane skeleton based on super-resolution microscopy results (Xu et al., 2013) exhibiting actin rings connected by spectrin tetramers. Ankyrin associated $Na_v$ channels anchor the lipid bilayer to the membrane skeleton. (B) A coarse-grain membrane skeleton dynamics model comprising representation of actin rings, spectrin filaments, and ankyrin. The insert shows the dimensions of the considered particles. (See also Figure S1.)

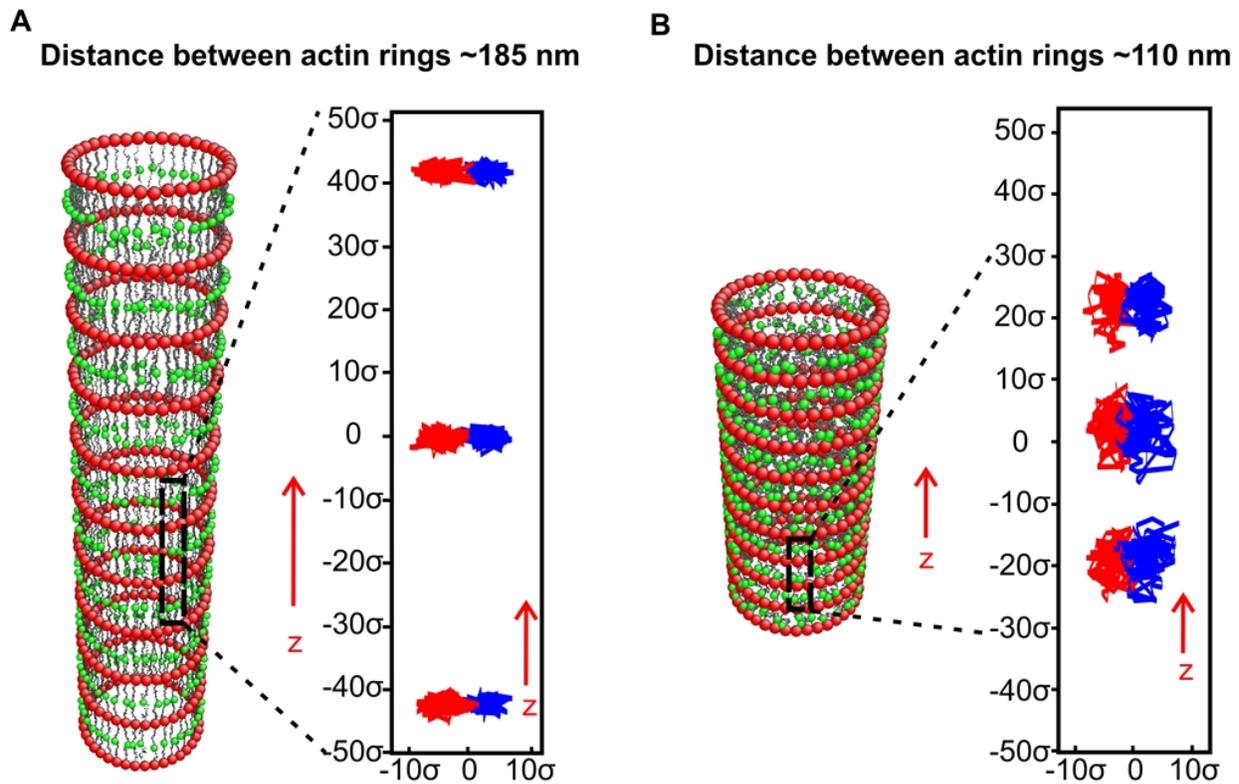

**Figure 2.** Membrane skeleton dynamics simulations. (A) The skeleton was equilibrated at a distance of approximately $185nm$ between actin rings while the trajectory of ankyrin particles (insert) was recorded for $5\times10^6$ time steps. (B) The skeleton was equilibrated at a distance of approximately $110nm$ between actin rings while the trajectory of ankyrin particles (insert) was recorded for $5\times10^6$ time steps. The longitudinal and circumferential separations of the trajectories of neighboring ankyrin particles, and consequently of the corresponding $Na_v$ channels, is well-defined in (A) but not in (B). (See also Figures S2 and S3.)

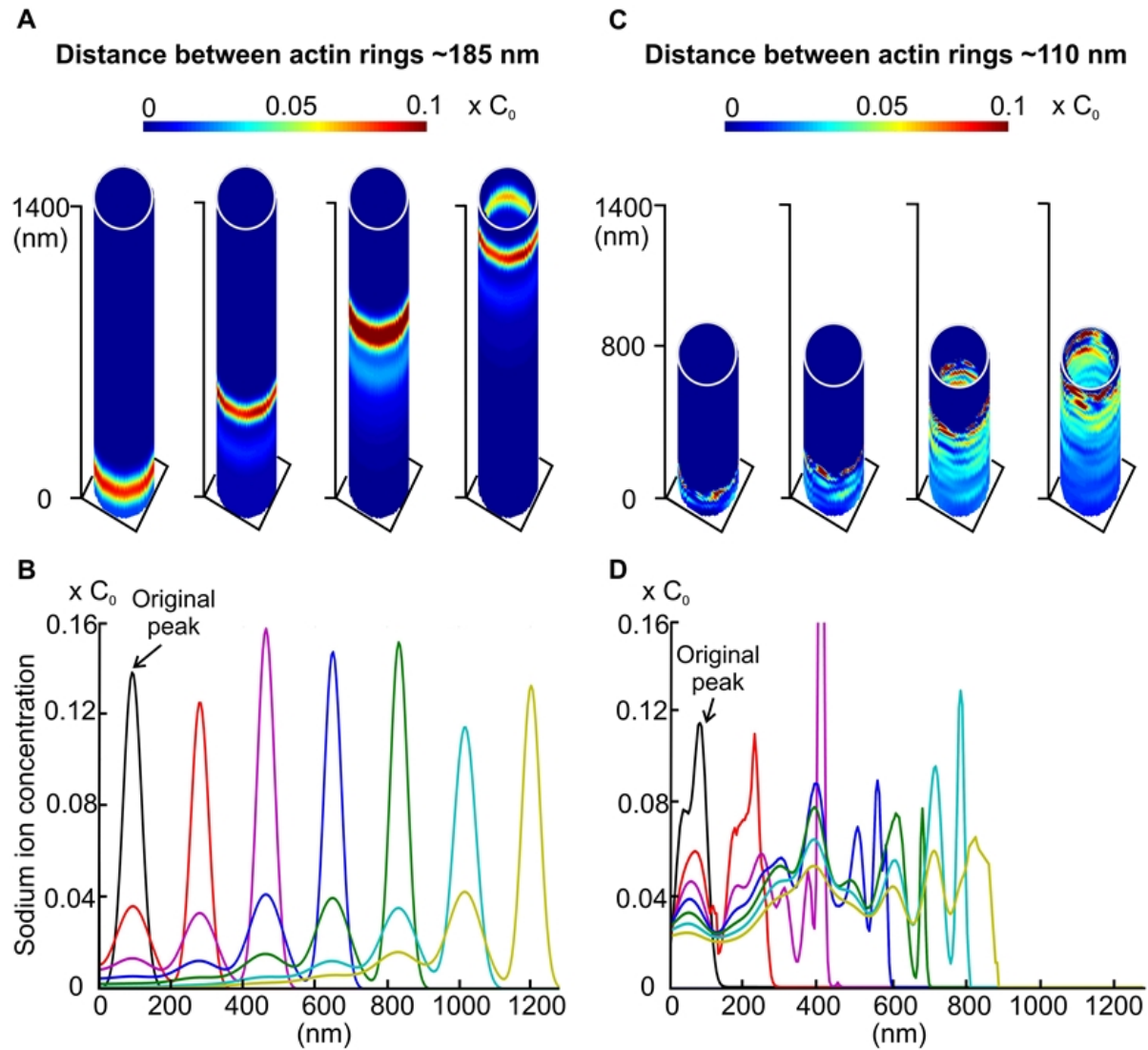

**Figure 3.** Coordinated propagation of sodium along the axon. An initial synchronous activation of Na$_v$ channels resulted in (A) the propagation of a band of sodium influx along the axon consistent with their ring-like arrangement and (B) the advancement of sodium influx down the axon with high fidelity and reproducibility. When spectrin filaments were not under entropic tension, the trajectories of neighboring Na$_v$ channels were not well-separated and an initial synchronous activation of Na$_v$ channels led to (C) the propagation of a sodium whirlpool and (D) the advancement of sodium influx down the axon with increased jitter and without reproducibility.

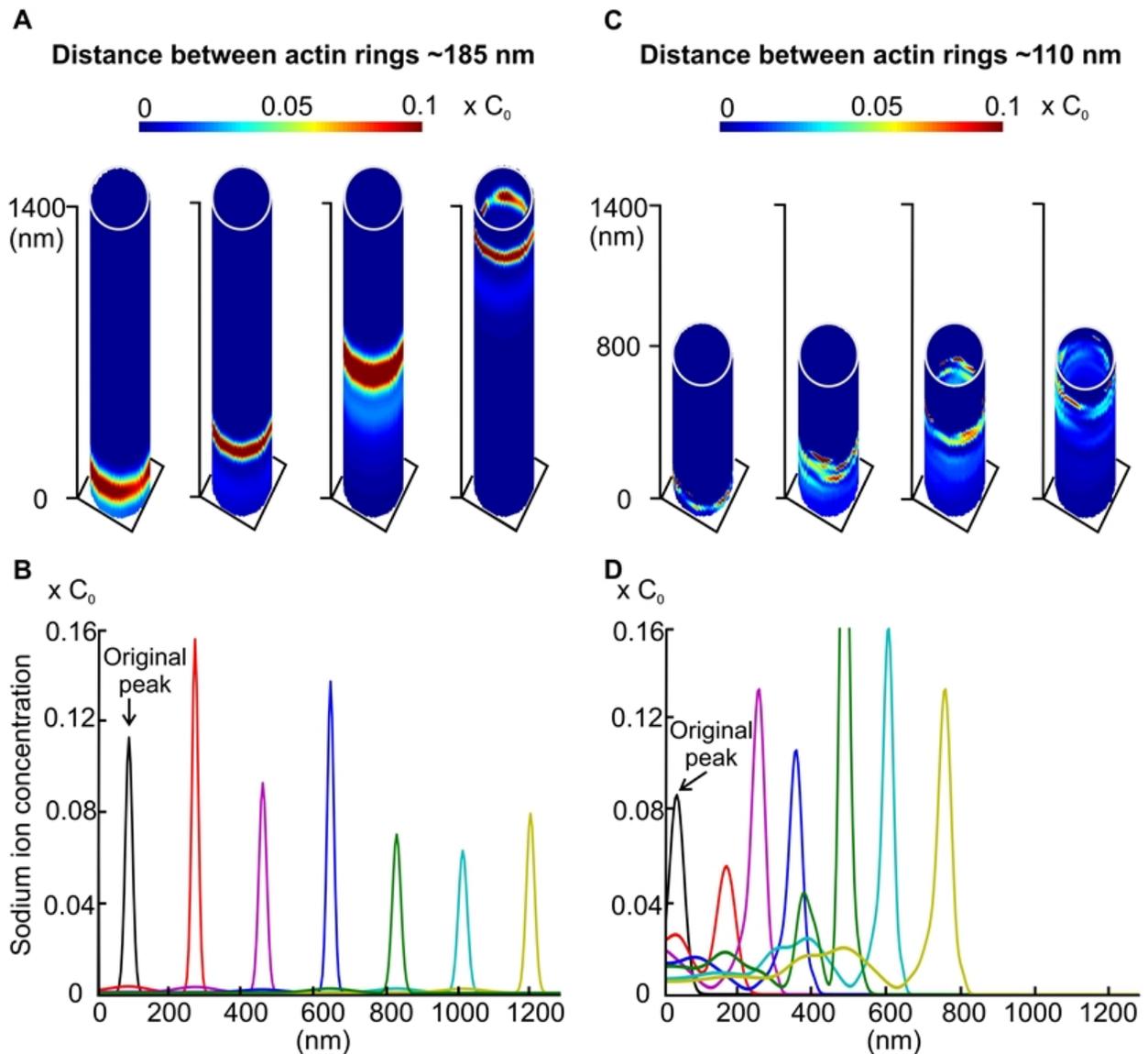

**Figure 4**. Coordinated propagation of sodium along the axon when the circumferential density of $Na_v$ channels was reduced to half. An initial synchronous activation of $Na_v$ channels resulted in (A) the propagation of a band of sodium influx along the axon consistent with their ring-like arrangement and (B) the advancement of sodium influx down the axon with high fidelity and reproducibility. When spectrin filaments were not under entropic tension, the trajectories of neighboring $Na_v$ channels were not well-separated and an initial synchronous activation of $Na_v$

channels led to (C) uncoordinated propagation of sodium and (D) the advancement of sodium influx down the axon with increased jitter and no reproducibility but at a lower degree than in the case of regular $Na_v$ circumferential distance.

**SUPPLEMENTAL INFORMATION**

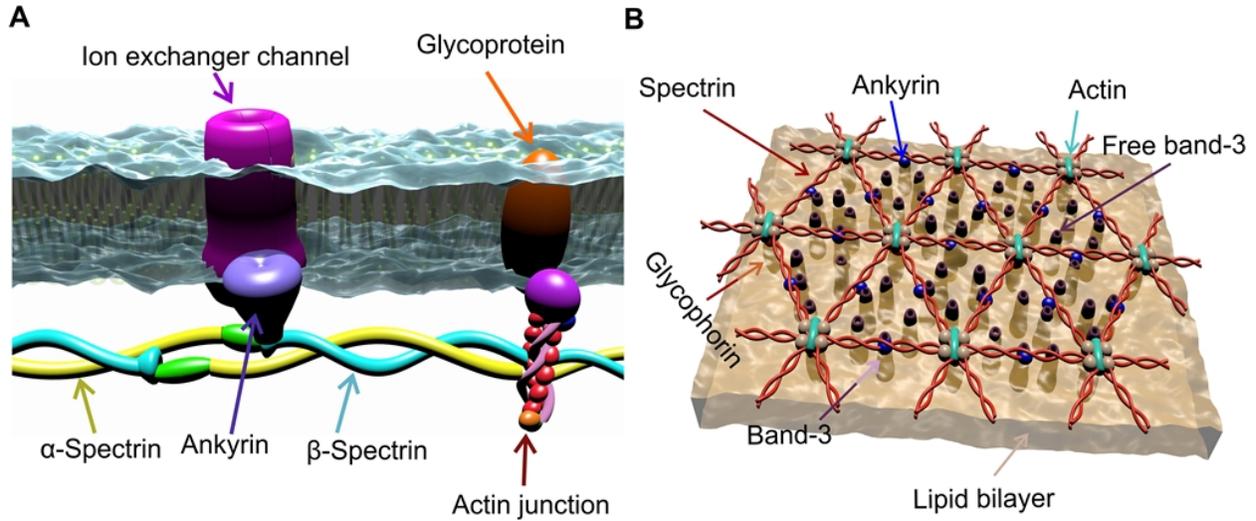

**Figure S1**. (A) Anchoring of the axon lipid bilayer to the membrane skeleton. Ankyrin binds to the 15$^{th}$ repeat of β-spectrin near its carboxyl terminus and to the anion exchanger band-3 in RBCs or to Na$_v$ in the axon. α-spectrin and β-spectrin filaments are connected at actin junctions. In RBCs the NH$_2$-terminal of β-spectrin binds to 4.1 which forms a membrane anchoring complex with glycophorin C (Bennett and Baines, 2001). In the axon however, it is not known how the lipid bilayer is anchored at the actin rings. Because of this, we indicate a generic glycoprotein. (B) Illustration of the RBC membrane skeleton comprising stretched spectrin tetramers connected at actin junctions and exhibiting a six-fold two-dimensional symmetry. The lipid bilayer is anchored to the membrane skeleton at actin junctions by glycophorin C and near the middle of spectrin tetramers by ankyrin which is connected to the anion exchanger band-3 protein (Alberts et al., 2008; Bennett and Baines, 2001).

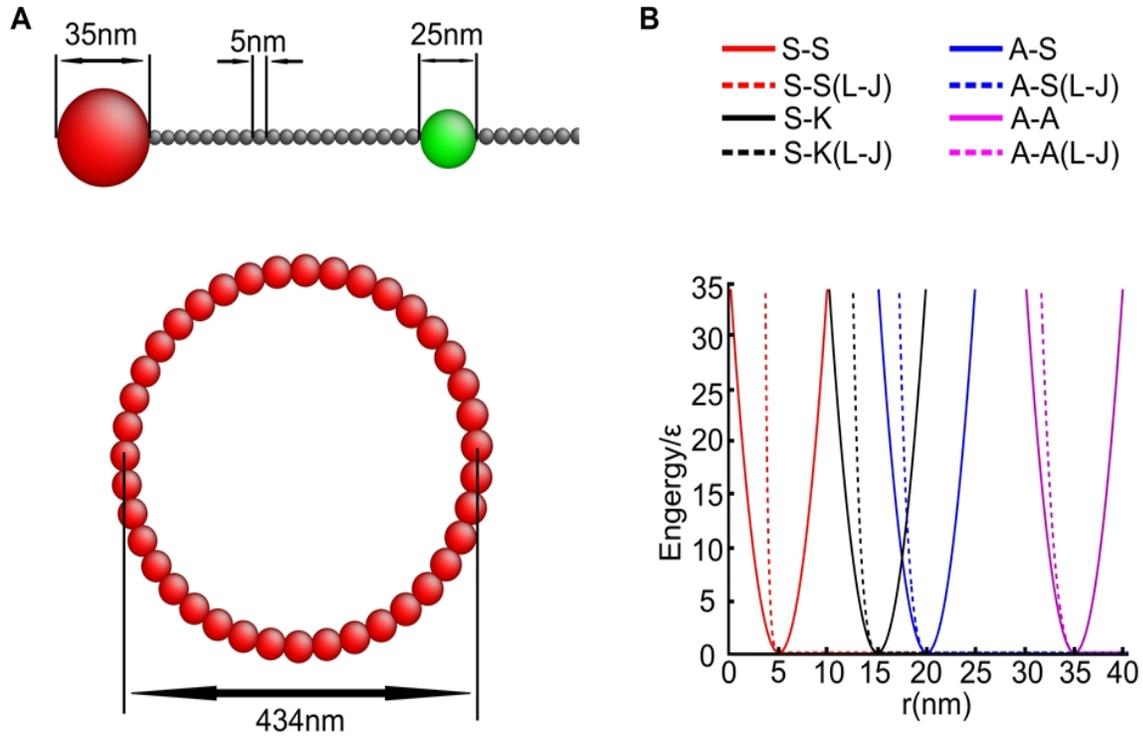

**Figure S2.** Detail of the axon membrane skeleton model. (A) Illustration of particles and connections involved in the model. Red particles represent actin junctions, gray particles represent spectrin subunits, and green particles represent ankyrin junctions. (B) The solid lines represent harmonic potentials applied between neighbor spectrin particle (S-S) of the same spectrin filament, ankyrin and spectrin (A-S), spectrin and ankyrin (A-K), and between neighboring actin particles (A-A) in the actin rings. Dashed lines represent Lennard-Jones (L-J) potential which represents steric repulsion between all particles used in the simulation.

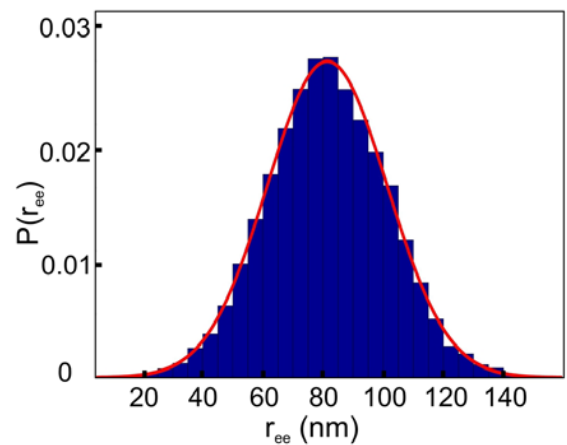

**Figure S3**. Histogram of the recorded end-to-end distances $(r_{ee})$ of a free spectrin filament during $3\times10^6$ time steps of a coarse-grain solvent-free molecular dynamics simulation at constant temperature. The associated normalized Gaussian probability density (red line) is also shown.